\documentclass[journal,twocolumn]{IEEEtran}
\usepackage{cite}
\usepackage{bm}
\usepackage{algorithm}
\usepackage{algcompatible}
\usepackage{float}
\usepackage{array}
\usepackage{multirow}
\usepackage[cmex10]{amsmath}
\usepackage{graphicx}
\usepackage{url}
\usepackage{color}
\usepackage{amssymb}
\usepackage{amsfonts}
\usepackage{amsmath}
\usepackage{extarrows}
\usepackage{graphicx}
\usepackage{amsfonts}
\usepackage{caption}
\usepackage{algorithm}
\usepackage{algpseudocode}
\usepackage{indentfirst}
\usepackage{caption}
\usepackage{esint} 
\usepackage{graphicx}
\usepackage{graphics}
\usepackage{pgfplots}
\usepackage{tikz}
\usepackage{epsfig}
\usepackage{soul}
\usepackage{caption}
\usepackage{gensymb}
\def\bA{{\mathbf{A}}} \def\bB{{\mathbf{B}}} \def\bC{{\mathbf{C}}} \def\bD{{\mathbf{D}}} 
 \def\bG{{\mathbf{G}}} \def\bH{{\mathbf{H}}} \def\bI{{\mathbf{I}}} \def\bJ{{\mathbf{J}}}
  \def\bM{{\mathbf{M}}}  
\def\bP{{\mathbf{P}}}    \def\bT{{\mathbf{T}}}
  \def\bW{{\mathbf{W}}}  

\def\ba{{\mathbf{a}}} \def\bb{{\mathbf{b}}} \def\bc{{\mathbf{c}}} \def\bd{{\mathbf{d}}} 
  \def\bh{{\mathbf{h}}}  
   \def\bn{{\mathbf{n}}} 
\def\bp{{\mathbf{p}}}    \def\bt{{\mathbf{t}}}
  \def\bw{{\mathbf{w}}}  \def\by{{\mathbf{y}}}

\begin{document}
\title{3D Localization with a Single Partially-Connected Receiving RIS: Positioning Error Analysis and Algorithmic Design
\author{Jiguang~He,~\IEEEmembership{Senior Member,~IEEE,} Aymen~Fakhreddine,~\IEEEmembership{Member,~IEEE,}\\ Charles~Vanwynsberghe,~\IEEEmembership{Member,~IEEE,} Henk Wymeersch,~\IEEEmembership{Senior Member,~IEEE,}\\ and George C. Alexandropoulos,~\IEEEmembership{Senior Member,~IEEE}}
\thanks{J. He, A. Fakhreddine, C. Vanwynsberghe, and G. C. Alexandropoulos are with the Technology Innovation Institute, 9639 Masdar City, Abu Dhabi, United Arab Emirates. G. C. Alexandropoulos is also with the Department of Informatics and Telecommunications, National and Kapodistrian University of Athens, 15784 Athens, Greece. H. Wymeersch is with the Department of Electrical Engineering, Chalmers University of Technology, 412 58 Gothenburg, Sweden. (e-mails: \{jiguang.he, aymen.fakhreddine, charles.vanwynsberghe\}@tii.ae, henkw@chalmers.se, alexandg@di.uoa.gr).}}
 \maketitle

\begin{abstract}
In this paper, we introduce the concept of partially-connected Receiving Reconfigurable Intelligent Surfaces (R-RISs), which refers to metasurfaces designed to efficiently sense electromagnetic waveforms impinging on them, and perform localization of the users emitting them. The presented R-RIS hardware architecture comprises subarrays of meta-atoms, with each of them incorporating a waveguide assigned to direct the waveforms reaching its meta-atoms to a reception Radio-Frequency (RF) chain, enabling signal/channel parameter estimation. We particularly focus on the scenarios where the user is located in the far-field of all the R-RIS subarrays, and present a three-Dimensional (3D) localization method which is based on narrowband signaling and Angle of Arrival (AoA) estimates of the impinging signals at each single-receive-RF R-RIS subarray. For the AoA estimation, which relies on spatially sampled versions of the received signals via each subarray's phase configuration of meta-atoms, we devise an off-grid atomic norm minimization approach, which is followed by subspace-based root MUltiple SIgnal Classification (MUSIC). The AoA estimates are finally combined via a least-squared line intersection method to obtain the position coordinates of a user emitting synchronized localization pilots. Our derived theoretical Cram\'er Rao Lower Bounds (CRLBs) on the estimation parameters, which are compared with extensive computer simulation results of our localization approach, verify the effectiveness of the proposed R-RIS-empowered 3D localization system, which is showcased to offer cm-level positioning accuracy. Our comprehensive performance evaluations also demonstrate the impact of various system parameters on the localization performance, namely the training overhead and the distance between the R-RIS and the user, as well as the spacing among the R-RIS's subarrays and its partitioning patterns.  
\end{abstract}

\section{Introduction}
Reconfigurable Intelligent Surfaces (RISs), mainly acting as passive yet tunable smart reflectors, have been recently introduced as a cost- and energy-efficient means to boost the performance of various wireless applications~\cite{Huang2018,Wu2019,huang2019holographic}, by altering the signal propagation environment~\cite{WavePropTCCN,Strinati2021Reconfigurable}. Among them belong the radio localization and sensing applications~\cite{wymeersch2019radio,He2019large,Elzanaty2021,he2019adaptive,Keykhosravi2022infeasible,Sundeep_RISISAC_2022}, especially in high-frequency systems, e.g., millimeter Wave (mmWave), where it has been identified that the deployment of RISs can create additional signal propagation paths, serving as extra degrees of freedom for the design of localization schemes. It has already been widely accepted \cite{CE_overview_2022} that RISs can provide a virtual Line-of-Sight (LoS) path from the Base Station (BS) to Mobile Stations (MSs) when the direct LoS path is spatiotemporally blocked; this situation can appear frequently in practice both in outdoor and indoor network deployments~\cite{RISE-6G-EUCNC,rise6g_scenarios}. Actually, even when the direct LoS path exists, an RIS can further enhance the localization performance by leveraging spatial diversity. In fact, an RIS can be regarded as an additional reference node (a.k.a. anchor) in addition to the BS. In certain envisioned applications, RISs can enable localization even when a BS or an access point is not part of the localization mechanism \cite{Keykhosravi2022infeasible,Alexandropoulos2022}. 

It has been shown in the literature that a single mmWave BS equipped with a large number of antenna elements and operating over a very large bandwidth can also achieve highly accurate positioning performance~\cite{Shahmansoori2017}. However, to accomplish this, the BS needs to rely on both temporal and angular channel parameters, e.g., time or time difference of arrival as well as Angles of Arrival (AoAs) or angles of departure. In principle, high-accuracy recovery of the angular channel parameters requires large-sized antenna arrays, and that of the temporal parameters requires large communications signal bandwidths. It is noted that, up to date, radio localization based on cellular network infrastructure is deemed as an additional functionality of BSs, implying that the network resources allocated to three-Dimensional (3D) localization  (especially bandwidth/spectrum and time) should be minimized so that the major network functionality, which is wireless communications, will not be significantly affected~\cite{wymeersch2019radio}. To this end, localization schemes that solely rely on angular channel parameters are highly desirable, since wideband operation is not necessary.
In this regard, in the recent work~\cite{Alexandropoulos2022}, multiple spatially distributed RISs, each equipped with a single Receive (RX) Radio Frequency (RF) and being capable to collect baseband measurements of spatially sampled versions of their impinging signal, were deployed to perform highly accurate 3D localization. However, the proposed framework necessitated wired/wireless backhaul links to gather the received signals from each RIS to a central processing unit, which was assigned with the localization computation task. This approach inevitably increases the deployment cost and the system's implementation complexity. 

In this paper, we capitalize on the localization framework of~\cite{Alexandropoulos2022} and present a partially-connected Receiving RIS (R-RIS) hardware architecture, comprising a few co-located single-RX-RF RIS subarrays, which can offer 3D localization in a computationally autonomous manner (i.e., a without the intervention of any BS or access point). Our intention is actually the development of a low-cost, low-complexity, energy-efficient, yet highly efficient, narrowband localization system, acting as the sole anchor for this computation task. In particular, the MS is assumed to transmit narrowband Sounding Reference Signals (SRSs), whose reception at the proposed partially-connected R-RIS, via its individual RX RF chains attached to distinct subarrays of meta-atoms, is used to obtain multiple independent angular measurements. These estimates are then mapped into the 3D Cartesian coordinates of the MS, which is assumed lying in the far-field of each R-RIS subarray. 

The proposed 3D localization system
requires only a small amount of bandwidth without any demand for wideband implementation. For the angular parameter estimation, we leverage the channel sparsity at mmWave frequencies, in the form of rank deficiency of its autocorrelation matrix, and use off-grid Atomic Norm Minimization (ANM)~\cite{Yang2015,Yang2016} in conjunction with subspace-based root MUltiple SIgnal Classification (MUSIC)~\cite{Barabell1983} to extract the AoAs of the LoS channel path at each R-RIS subarray. Thanks to the similar path loss and received Signal-to-Noise Ratio (SNR), all the angular estimates possess similar performance in terms of the estimation error variance. Thus, for the mapping from the AoA estimates to the 3D coordinate of the MS, a Least Squares (LS) approach\footnote{In general, if the error variance for the estimation at each R-RIS subarray differs, a weighted LS will offer better performance compared to its unweighted counterpart. However, in all investigated cases in this paper, the error variances for the AoA estimates are approximately equal, thus, the weight matrix is in the form of a scaled identity matrix. In this sense, the weighted LS offers the same solution with the classical LS.} is sufficient. Our derived theoretical Cram\'er Rao Lower Bounds (CRLBs) and simulated performance evaluation results showcase that the proposed R-RIS-based single-anchor 3D localization system can achieve cm-level estimation accuracy. In addition, we investigate the impact of the training overhead, MS-to-R-RIS distance, inter R-RIS subarray spacing, and the R-RIS partitioning patterns on the localization performance, offering useful insights for the practical deployment of the proposed system. The contributions of this paper are summarized as follows:
\begin{itemize}
    \item We present a partially-connected R-RIS hardware architecture and study the feasibility of its deployment for highly accurate 3D localization from both theoretical and algorithmic perspectives. Our extensive numerical investigations showcase that the proposed single-anchor system can achieve cm-level localization in a cost- and time-efficient manner. 
    \item We analyze the influence of the existence of Non-Line-of-Sight (NLoS) paths on the angular parameter estimation of the LoS path, considering two different assumptions for the angle difference among them, specifically, a deterministic and a random difference. 
    \item We present a localization algorithm, which is based on off-grid ANM for high-precision channel vector estimation, followed by subspace-oriented root MUSIC intended for super-resolution recovery of the angular parameters. The proposed algorithm approaches the theoretical performance limits, as characterized by our CRLB analyses.
    \item We present a method for mapping the estimate of the angular channel parameters for the MS to its 3D coordinates, which includes an outliers finding method that excludes the respective estimates to eliminate their negative effect on the localization performance. 
    \item We provide comprehensive performance evaluations of the role of key system parameters (i.e., training overhead, MS-to-R-RIS distance, inter R-RIS subarray spacing, and the R-RIS partitioning patterns) on the accuracy of the presented localization framework.
\end{itemize}

The remainder of the paper is organized as follows: Section~II introduces the proposed 3D localization system model as well as the channel model, the geometric relationships of its parameters, and the received signal model. Section~III presents the theoretical performance limits via CRLB analyses, where the impact of the angle difference between the LoS and NLoS paths is investigated. The proposed 3D localization algorithm, leveraging ANM and subspace-oriented root MUSIC, and performing close to our derived CRLBs, is provided in Section~IV. Our numerical evaluations on the localization performance are presented in Section~V, including extensive investigations over various system operation parameters (training overhead, MS-to-R-RIS distance, inter R-RIS subarray spacing, and R-RIS partitioning patterns). Finally, the paper is concluded and some direction for potential future works are presented in Section~VI.      

\textit{Notations}: Bold lowercase letters denotes vectors (e.g., $\ba$), while bold capital letters represent matrices (e.g., $\bA$). The operators $(\cdot)^*$, $(\cdot)^\mathsf{T}$, $(\cdot)^\mathsf{H}$, and $(\cdot)^{-1}$ denote conjugate of a complex number, the matrix or vector transpose, Hermitian transpose, and matrix inverse, respectively. $\mathrm{diag}(\ba)$ denotes a square diagonal matrix with the entries of $\ba$ on its maindiagonal, $\otimes$ denotes the Kronecker product, 
$\mathbb{E}[\cdot]$ is the expectation operator, $\mathbf{0}$ denotes the all-zero vector or matrix, $\bI_{M}$ ($M\geq2$) denotes the $M\times M$ identity matrix, and $j = \sqrt{-1}$. $\mathrm{Tr}(\cdot)$ and $\mathrm{Toep}(\cdot)$ represent the trace operator and a Toeplitz matrix formulated by the argument within the brackets, respectively, $\|\cdot\|_2$ denotes the Euclidean norm of a vector, and $|\cdot|$ returns the absolute value of a complex number. $[\ba]_m$, $[\bA]_{mn}$, $[\ba]_{m:n}$, and $[\bA]_{m:n, m:n}$ denote the $m$th element of $\ba$, the $(m,n)$th element or submatrix of $\bA$, the subvector of $\ba$ composed of its elements with indices $m,m+1, \ldots, n$, and the submatrix of $\bA$ formed by its elements in the rows $m,m+1, \ldots, n$ and columns $m,m+1, \ldots, n$. $\mathcal{U}[a,\;b]$ stands for the uniform distribution within the range $[a,\;b]$ and $\mathcal{CN}(a,b)$ denotes the complex Gaussian distribution with mean $a$ and variance $b$. Finally, $\Re\{\cdot\}$ returns the real part of its complex argument.

\section{System and Channel Models}\label{sec_Sys_Model}
In this section, we introduce the proposed 3D localization system model, which is based on a single partially-connected R-RIS composed of a few single-RX-RF subarrays of meta-atoms. The considered channel model, the geometric relationships of its parameters that are essential for position estimation, and the received signal model are also presented.   
\begin{figure}[t]
	\centering
\includegraphics[width=1 \linewidth]{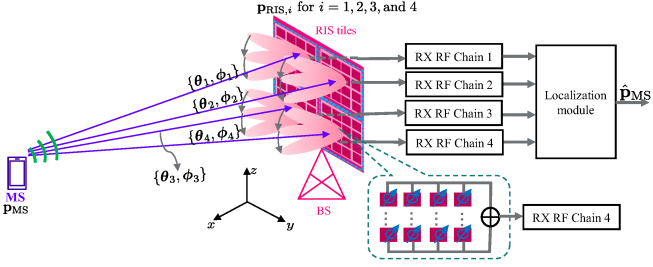}
	\caption{The proposed 3D localization system model which is based on a single partially-connected R-RIS architecture. The metasurface is partitioned into subarrays of meta-atoms, with each subarray attached to an RX RF chain, and processes pilots intended for localization. When a MS transmits the localization SRSs, the R-RIS computes the user's 3D position via processing the baseband received signals at the outputs of the RX RF chains of its distinct subarrays. 
 }
		\label{System_Model}
		\vspace{-0.5cm}
\end{figure}

\subsection{Proposed Localization System}
The proposed 3D localization system, which is based on a single partially-connected R-RIS hardware architecture comprising subarrays of meta-atoms and autonomously estimates the position of mobile users lying within the surface's area of influence \cite{rise6g_scenarios}, is illustrated in Fig.~\ref{System_Model}.\footnote{Although Fig.~\ref{System_Model} depicts an R-RIS as a uniform planar array with regular partitioning comprising four subarrays, our proposed AoA-based 3D localization system requires, in principle, only three subarrays. The impact of the number of subarrays and that of the R-RIS partitioning on the localization accuracy of our localization scheme will be investigated in Section~\ref{Numerical_results}. It is noted that the proposed RIS framework comprising subarrays of meta-atoms could replace the RF front-end of a conventional multi-antenna BS, realizing a metasurface-based holographic RX \cite{9324910,huang2020holographic} capable of MS localization through narrowband measurements.
} Following the metasurface hardware designs in \cite{hardware2020icassp,9324910,HRIS_Nature}, each subarray incorporates a waveguide that directs the impinging signals on its meta-atoms to a single output port feeding a reception RF chain. Note that such an RX RF chain consists of a low noise amplifier, a mixer that downconverts the signal from RF to baseband, and an analog-to-digital converter \cite{FD_MIMO_Arch}. The phase configurations of all subarray's elements result effectively in an analog beamformer/combiner realizing a spatial reception filter, which is applied on the impinging waveform. In this paper's framework, this feature implies the spatial sampling of the MS transmitted pilots intended for localization. We assume that the MS is located far from each R-RIS subarray.  Without loss of generality, we further assume a single-antenna MS, and leave the multi-antenna case as well as the tracking problem for the case of mobile MSs for future works.

We consider that the R-RIS consists of single-RX-RF Uniform Planar subArrays (UPAs) of meta-atoms and, according to Fig.~\ref{System_Model}, is placed parallel to the $y\text{-} z$ plane. It is also shown in the figure that the outputs of the reception RF chains of all subarrays feed a baseband AoA estimation and localization module, which is part of the R-RIS controller and is based on the extraction of the impinging signal's angular parameters with respect to each subarray, similar to the localization mechanism of multi-anchor systems operating in narrowband transmission conditions~\cite{Alexandropoulos2022}. To this end, the R-RIS controller has basic storage and computing capabilities.\footnote{It is noted that the proposed baseband AoA estimation and localization module can be a part of the R-RIS. In this case, the R-RIS controller collecting the output signals from the reception RF chains needs to convey them to the AoA estimation and localization module.} 
The impact of the R-RIS partitioning on the position estimation, whose optimization is left for future work, will be investigated in Section~\ref{subsec_RIS_partitioning}.

\subsection{Channel Model}
In this paper, we consider the general channel model used in~\cite{ozturk2022ris} for formulating the channel gain vector $\bh\in\mathbb{C}^{M}$ for the wireless link between the MS and the R-RIS, which is:
\begin{equation}\label{h_MS_RIS}
    \bh \triangleq \sum_{l = 1}^{L}  \ba_l /\sqrt{\rho_{l}},
\end{equation}
where $M$ and $L$ represent the numbers of the R-RIS elements and channel paths, whereas $\rho_l$ and $\ba_l \in \mathbb{C}^M$ are the path loss and near-field R-RIS steering vector associated with the $l$th path, respectively. Without loss of generality, we assume in the sequel that $l =1$ is associated with the LoS path in $\bh$, and the rest are associated with the NLoS paths. Taking the LoS path as an example, the $m$th coefficient of $\ba_1$ can be expressed as 
\begin{equation}
    [\ba_1]_m \triangleq \exp\Big( j \frac{2 \pi}{\lambda} \big(\|\bp_{\text{MS}} - \bp^{(m)}_{\text{RIS}}\| -\|\bp_{\text{MS}} - \bp_{\text{RIS},c}\| \big)\Big),
\end{equation}
where $\bp_\text{MS} \triangleq [x_\text{MS},y_\text{MS},z_\text{MS}]^\mathsf{T}$, $\bp^{(m)}_{\text{RIS}}$, and $\bp_{\text{RIS},c}$ are the 3D Cartesian coordinates of the MS, the $m$th R-RIS element, and the centroid of the R-RIS, respectively, and $\lambda$ is the wavelength. By following ~\eqref{h_MS_RIS}, the channel vector $\bh_i \in \mathbb{C}^{M_i}$ between the MS and each $i$th R-RIS subarray, where $i=1,2,\ldots,I$ (an example with $I =4$ R-RIS subarrays is included in Fig.~\ref{System_Model}), is mathematically expressed as $\bh_i \triangleq \sum_{l = 1}^{L_i}  \ba_{i,l} /\sqrt{\rho_{i,l}}$, where $M_i$ and $L_i$ represent the number of elements for the $i$th R-RIS subarray and the total number of paths in $\bh_i$, respectively, and $ \rho_{i,l}$ is the corresponding path loss. Similarly, $[\ba_{i,1}]_m$ with $l=1$ is in the following form: 
\begin{equation}\label{NF_steering_vector}
    [\ba_{i,1}]_m \triangleq \exp\Big( j \frac{2 \pi}{\lambda} \big(\|\bp_{\text{MS}} - \bp^{(m)}_{\text{RIS},i}\| -\|\bp_{\text{MS}} - \bp_{\text{RIS},i}\| \big)\Big), 
\end{equation}
where $ \bp^{(m)}_{\text{RIS},i}$ and $\bp_{\text{RIS},i}$ are the 3D Cartesian coordinates of the $m$th element and the centroid of the $i$th R-RIS subarray, respectively.
When the MS is located far from the R-RIS subarrays, i.e.,  $\|\bp_{\text{MS}} - \bp_{\text{RIS},i}\| \gg \|\bp_{\text{MS}} - \bp^{(m)}_{\text{RIS},i}\|$, the steering vector $\ba_{i,1}$ in $\bh_i$ can be approximated as~\cite{ozturk2022ris}: 
$\ba_{i,1} \approx e^{-j 2\pi  d_{i,1}/\lambda} \boldsymbol{\alpha}_y(\theta_{i,1},\phi_{i,1}) \otimes \boldsymbol{\alpha}_z(\phi_{i,1})$ with $ d_{i,1} = \|\bp_{\text{MS}} - \bp_{\text{RIS},i}\|$ and $\theta_{i,1}$ and $\phi_{i,1}$ being the LoS azimuth and elevation AoAs at the $i$th R-RIS subarray; this approximation can be used for the NLoS paths as well. The mathematical expressions for the array response vectors  $\boldsymbol{\alpha}_y(\theta_{i,1},\phi_{i,1}) \in \mathbb{C}^{M_{i,y}}$ and $\boldsymbol{\alpha}_z(\phi_{i,1})\in \mathbb{C}^{M_{i,z}}$ in the previous approximation, where $M_i \triangleq M_{i,y} M_{i,z}$ with $M_{i,y}$ and $M_{i,z}$ being the number of the horizontal and vertical elements of the $i$th R-RIS subarray, are given as follows:
\begin{align}\label{alpha_y}
    \boldsymbol{\alpha}_y(\theta_{i,1},\phi_{i,1}) \triangleq& \Big[1, e^{j \frac{2\pi  d_{i,y}}{\lambda}  \sin(\theta_{i,1}) \sin(\phi_{i,1})}, \nonumber\\
    & \cdots, e^{j \frac{2\pi d_{i,y}}{\lambda} (M_{i,y} -1) \sin(\theta_{i,1})\sin(\phi_{i,1})} \Big]^{\mathsf{T}},\\
       \boldsymbol{\alpha}_z(\phi_{i,1}) \triangleq& \Big[1, e^{j \frac{2\pi d_{i,z}}{\lambda}  \cos(\phi_{i,1}) }, \nonumber\\
    & \cdots, e^{j \frac{2\pi d_{i,z}}{\lambda} (M_{i,z} -1) \cos(\phi_{i,l})} \Big]^{\mathsf{T}},
 \end{align}
where $d_{i,y}$ and $d_{i,z}$ are the inter-element spacings for the horizontal and vertical elements in each $i$th R-RIS subarray. It is noted that, for quasi-free-space propagation conditions, fine-grained control over the reflected signals from each R-RIS subarray is essential for accurate reflective beamforming. This fact motivated researchers to rely on meta-atoms of sub-wavelength sizes \cite{huang2020holographic}, despite inevitable strong mutual coupling among those elements \cite{alexandg_ESPARs,Gradoni2020} (e.g., when the spacing of adjacent meta-atoms is $\lambda/10$). In contrast, in rich scattering environments, the wave energy is statistically equally spread throughout the wireless medium, and $\lambda/2$-sized meta-atoms, and consequently inter-element spacing, suffice \cite{alexandg_2021}.   

Recall that we assume that MS is located far from each R-RIS subarray. Based on this assumption, $\bh_i$ can be reformulated as 
\begin{equation}\label{h_i}
    \bh_i \triangleq \sum_{l = 1}^{L_i}\frac{e^{-j 2\pi  d_{i,l}/\lambda}}{\sqrt{\rho_{i,l}}} \boldsymbol{\alpha}_y(\theta_{i,l},\phi_{i,l}) \otimes \boldsymbol{\alpha}_z(\phi_{i,l}),
\end{equation}
where $d_{i,l}$ denotes the (total) Euclidean distance between the MS and the $i$th R-RIS subarray over the $l$th path, $\theta_{i,l}$ and $\phi_{i,l}$ are the $l$th azimuth and elevation AoAs at the $i$th R-RIS subarray. We consider the free-space path loss for $\rho_{i,1}$, which is modeled as~\cite{Rappaport2013} $\rho_{i,1} =10^{3.245} d_{i,1}^2 f_c^2$
with $f_c$ (in GHz) being the carrier frequency and defined as $f_c = \frac{ c }{\lambda}$, where $c$ is the speed of light. In Fig.~\ref{System_Model}, we define the angle vectors $\boldsymbol{\theta}_i \triangleq[\theta_{i,1}\,\cdots\,\theta_{i,L_i}]^\mathsf{T} \in \mathbb{R}^{L_i}$ and $\boldsymbol{\phi}_i \triangleq[\phi_{i,1}\,\cdots\, \phi_{i,L_i}]^\mathsf{T}\in \mathbb{R}^{L_i}$ for the collection of the same type of angles. To model the impact on the some key system parameters of the proposed co-located deployment of the $I$ R-RIS subarrays, we assume that: \textit{i}) $d_{1,1} \approx d_{2,1} \approx \cdots \approx d_{I,1}$; \textit{ii}) $\theta_{1,1} \approx \theta_{2,1} \approx \cdots \approx \theta_{I,1}$; and \textit{iii}) $\phi_{1,1} \approx \phi_{2,1} \approx \cdots \approx \phi_{I,1}$. We also make the assumption that $\rho_{i,1} \ll \rho_{i,2} \leq \cdots \leq \rho_{i,L_i}$.


The geometric relationship between the proposed R-RIS-based 3D localization system and the MS to be localized is essential for the location estimation. 
By introducing the three-tuple vector $\boldsymbol{\xi}_i \triangleq [\cos(\theta_{i,1}) \cos(\phi_{i,1}), \sin(\theta_{i,1}) \cos(\phi_{i,1}), \sin(\phi_{i,1}) ]^\mathsf{T}$ including the azimuth and elevation AoAs, the geometric relationship between the MS and the $i$th R-RIS subarray is:
\begin{equation}\label{Geometry}
    \bp_{\text{MS}} = \bp_{\text{RIS},i} + d_{i,1} \boldsymbol{\xi}_i.
\end{equation}

\subsection{Received Signal Model}
A sequence of SRSs are transmitted over the uplink from the MS to the R-RIS with several subarrays. The received signal at each $i$th R-RIS subarray during the $k$th time slot can be expressed as follows:
\begin{equation}
    y_{i,k} \triangleq \bw_{i,k}^\mathsf{H} \bh_i s_{k} + n_{i,k},
\end{equation}
where $s_{k} \in \mathbb{C}$ is the SRS with the power constraint $ \mathbb{E}[ |s_{k}|^2] = P$, and $\bw_{i,k} \in \mathbb{C}^{M_i}$ is the effective analog combining vector at the $i$th R-RIS subarray during the $k$th time slot, such that $|[\bw_{i,k}]_m|=1$ $\forall$$m = 1,2,\ldots,M_i$. In addition, $n_{i,k} \in \mathbb{C}$ follows $\mathcal{CN}(0,\sigma_i^2)$. By collecting the received signals across a total of $K$ time slots (implying the training overhead) to the vector $\by_i \triangleq [y_{i,1}\,\cdots\, y_{i,K} ]^\mathsf{T} \in \mathbb{C}^{K}$, and setting $s_{k}$ to $\sqrt{P}$ $\forall k$ for the sake of simplicity, the following expression is deduced:
\begin{equation} \label{by_i}
    \by_i = \sqrt{P}\bW_i^\mathsf{H} \bh_i + \bn_i,
\end{equation}
where $\bW_i \triangleq [\bw_{i,1}\,\cdots\,\bw_{i,K} ]\in \mathbb{C}^{M_i \times K}$ and $\bn_i \triangleq [n_{i,1}\,\cdots\, n_{i,K} ]^\mathsf{T}\in \mathbb{C}^{K}$.

By using the received signal vector $\by_i$, the $i$th R-RIS subarray can extract the angular parameters in the LoS path of $\bh_i$, i.e., $\theta_{i,1}$ and $\phi_{i,1}$. Then, the mapping of the estimated angles at all R-RIS subarrays to the 3D coordinate of the MS can be performed based on the pre-known coordinates of the R-RIS subarrays, as will be shown in the sequel.

\section{CRLB Analyses}\label{CRLB_Analyses}
In this section, by following the CRLB analysis~\cite{kay1993fundamentals}, we present mathematical derivations for the theoretically achievable performance limits of 3D localization with the proposed system model. We also investigate analytically the role of MultiPath Components (MPCs), and in particular the angle difference between the LoS and NLoS paths, on the angular parameter estimation problem for the LoS path, which in turn impacts 3D localization.

\subsection{CRLB on the Channel Parameters Estimation} 
By defining $\boldsymbol{\mu}_i \triangleq \bW_i^\mathsf{H} \bh_i \in \mathbb{C}^{K}$ focusing on each $i$th R-RIS subarray, we derive the Fisher Information Matrix (FIM) according to the partial derivatives related to the channel parameters ($\theta_{i,l}$, $\phi_{i,l}$, and $d_{i,l}$), which can be computed as follows:
\begin{align}\label{mu_over_theta}
    \frac{\partial \boldsymbol{\mu}_i} {\partial \theta_{i,l}} &= \bW_i^\mathsf{H} \Big(\bD_{i,l}^{(1)} \otimes \bI_{M_{i,z}}\Big)\bh_{i,l} ,\\
     \frac{\partial \boldsymbol{\mu}_i} {\partial \phi_{i,l}} &=  \bW_i^\mathsf{H} \Big(\bD_{i,l}^{(2)} \otimes \bI_{M_{i,z}}  + \bI_{M_{i,y}} \otimes \bD_{i,l}^{(3)}\Big)\bh_{i,l}, \\
    \frac{\partial \boldsymbol{\mu}_i} {\partial d_{i,l}} & =  \frac{-j\pi-1}{d_{i,l}}\bW_i^\mathsf{H}  \bh_{i,l},\label{mu_over_d}
\end{align}
where $\bh_{i,l} \triangleq\frac{e^{-j 2\pi  d_{i,l}/\lambda}}{\sqrt{\rho_{i,l}}} \boldsymbol{\alpha}_y(\theta_{i,l},\phi_{i,l}) \otimes \boldsymbol{\alpha}_z(\phi_{i,l})$ is the $l$th channel term in $\bh_i$. The other newly introduced notations are 
\begin{align}
\bD_{i,l}^{(1)} = &\;\mathrm{diag}\Big(\Big[0,j\pi \cos(\theta_{i,l}) \sin(\phi_{i,l}), \cdots, \nonumber\\ &j\pi \Big(M_{i,y} -1\Big) \cos(\theta_{i,l}) \sin(\phi_{i,l})\Big]\Big),\\
\bD_{i,l}^{(2)} = &\;\mathrm{diag}\Big(\Big[0,j\pi  \sin(\theta_{i,l}) \cos(\phi_{i,l}), \cdots,\nonumber\\ &j\pi \Big(M_{i,y} -1\Big) \sin(\theta_{i,l}) \cos(\phi_{i,l})\Big]\Big),\\
\bD_{i,l}^{(3)} = &\;\mathrm{diag}\Big(\Big[0,-j\pi \sin(\phi_{i,l}) \cdots,\nonumber\\ &-j\pi \Big(M_{i,z} -1\Big)  \sin(\phi_{i,l})\Big]\Big).
\end{align}
The $(m,n)$th element of the FIM $\bJ(\boldsymbol{\nu}_i) \in \mathbb{R}^{3L_i \times 3 L_i}$ associated with the $3L_i$-tuple channel parameters $\boldsymbol{\nu}_i \triangleq [\theta_{i,1}, \phi_{i,1}, d_{i,1},\cdots, \theta_{i,L_i}, \phi_{i,L_i}, d_{i,L_i}]^\mathsf{T}$ can be calculated as
\begin{equation}
[\bJ(\boldsymbol{\nu}_i) ]_{mn} =  \frac{2P}{\sigma_i^2}\Re \Big\{\frac{\partial \boldsymbol{\mu}_i^\mathsf{H}} {\partial [\boldsymbol{\nu}_i]_m} \frac{ \partial \boldsymbol{\mu}_i}{ \partial [\boldsymbol{\nu}_i]_n} \Big\},
\end{equation}
where the partial derivatives in~\eqref{mu_over_theta}--\eqref{mu_over_d} need to be substituted.

The error covariance matrix of $\boldsymbol{\nu}_i$ is then bounded as 
\begin{equation}\label{CRLB_ang_par}
    \mathbb{E}\Big\{(\boldsymbol{\nu}_i -\hat{\boldsymbol{\nu}}_i )(\boldsymbol{\nu}_i -\hat{\boldsymbol{\nu}}_i )^\mathsf{T}\Big\} \geq \bJ^{-1}(\boldsymbol{\nu}_i),
\end{equation}
where $\hat{\boldsymbol{\nu}}_i$ is supposed to be the unbiased estimate of $\boldsymbol{\nu}_i$. In order to guarantee the non-singularity of $\bJ(\boldsymbol{\nu}_i)$ in~\eqref{CRLB_ang_par}, the training overhead should meet the following necessary condition: $K \geq 3 L_i$. 

\subsection{Effect of NLoS Paths on LoS Angular Parameter Estimation}\label{Effect_of_NLoS_Paths}
There are existing works that exploit both LoS and NLoS paths for localization, e.g., ~\cite{Witrisal2016, Shahmansoori2017, Mendrzik2019}. However, in this paper, for the sake of tractability/simplicity, we only focus on extracting the angular parameters of the LoS path for the 3D localization. It will be shown in the results (see Section~\ref{Numerical_results}) that our approach can achieve high-precision localization performance up to cm level. By following this principle, let us define the matrices $\bG_{i,1} \triangleq [\frac{\partial \boldsymbol{\mu}_i} {\partial \theta_{i,1}}, \frac{\partial \boldsymbol{\mu}_i} {\partial \phi_{i,1}}] \in \mathbb{C}^{K \times 2}$ and $\bG_{i,2} \triangleq [\frac{\partial \boldsymbol{\mu}_i} {\partial d_{i,1}}, \frac{\partial \boldsymbol{\mu}_i} {\partial \theta_{i,2}}, \frac{\partial \boldsymbol{\mu}_i} {\partial \phi_{i,2}}, \frac{\partial \boldsymbol{\mu}_i} {\partial d_{i,2}}, \cdots,\frac{\partial \boldsymbol{\mu}_i} {\partial \theta_{i,L_i}}, \frac{\partial \boldsymbol{\mu}_i} {\partial \phi_{i,L_i}}, \frac{\partial \boldsymbol{\mu}_i} {\partial d_{i,L_i}}  ] \in \mathbb{C}^{K \times (3 L_i -2)}$, which are the collections of the partial derivatives related to the angular parameters in the LoS path of $\bh_i$ and that of the remaining ones, respectively. Therefore, $\bJ^{-1}(\boldsymbol{\nu}_i)$ in~\eqref{CRLB_ang_par} can be presented in the form of~\cite{scharf1993geometry,Pakrooh2015}:
\begin{align}\label{bJ_inv_nu_mul_path}
      &\bJ^{-1}(\boldsymbol{\nu}_i)= \nonumber \\
      &  \frac{\sigma_i^2}{2P}\!\!
      \begin{bmatrix}
     \Big(\bG_{i,1}^\mathsf{H} (\bI_{K} \!-\! \bP_{\bG_{i,2}})\bG_{i,1}\Big)^{-1} & * \\
      * &  \!\!\!\!\!\!\Big(\bG_{i,2}^\mathsf{H} (\bI_{K} \!-\! \bP_{\bG_{i,1}})\bG_{i,2} \Big)^{-1}
      \end{bmatrix}\!\!,
\end{align}
where $\bP_{\bG_{i,m}} = \bG_{i,m}( \bG_{i,m}^\mathsf{H} \bG_{i,m})^{-1}\bG_{i,m}^\mathsf{H} $ is the orthogonal projection onto the column space of $\bG_{i,m}$ for $m = 1$ and $2$. 

By inspecting~\eqref{bJ_inv_nu_mul_path}, we next study the effect of $\bG_{i,2}$ on the first submatrix on the diagonal of $\bJ^{-1}(\boldsymbol{\nu}_i)$, i.e., $\frac{\sigma_i^2}{2P}(\bG_{i,1}^\mathsf{H} (\bI_{K} - \bP_{\bG_{i,2}})\bG_{i,1})^{-1}$, which represents the estimation error bound of $[\theta_{i,1},\phi_{i,1}]^\mathsf{T}$. It is readily known that this influence comes from the orthogonal projection term $\bP_{\bG_{i,2}}$. Two examples will be provided in the sequel for different assumptions on the angle difference between the LoS and NLoS paths. 

\begin{figure}[t]
	\centering
\includegraphics[width=0.99\linewidth]{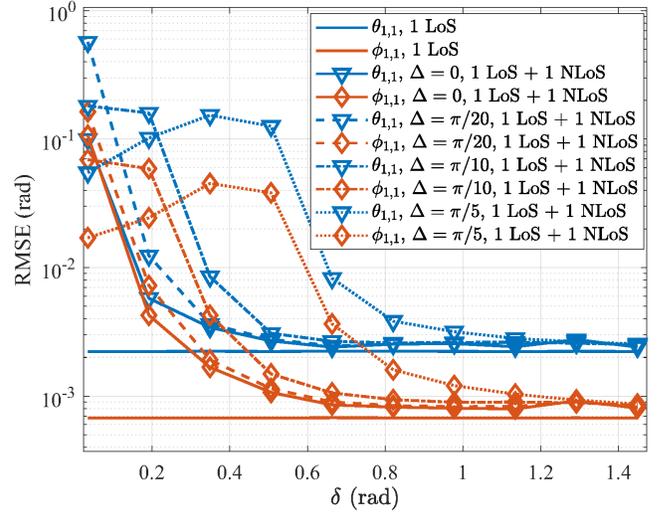}
	\caption{The effect of angle difference $\delta$ between the LoS path and a sole NLoS path on the angular parameter estimation of the former, where the power ratio between the LoS and the NLoS paths is set to be $20$ dB. We have set $K = 64$, $M_1 = 4 \times 4$, and $P = 0$ dBm. }
		\label{Effect_of_NLoS_on_Ang_Est}
\end{figure}
\subsubsection{Example 1} In this study, we take the first R-RIS subarray as an example (i.e., $i =1$) and evaluate the angular parameter estimation performance as a function of a deterministic and fixed angle difference $\delta$ between the LoS path and a sole NLoS path. We consider the scenario of a two-path channel and set the angle difference $\delta$ the same for both azimuth and elevation AoAs for the ease of tractability, i.e., $\theta_{1,2} = \theta_{1,1} + \delta$ and $\phi_{1,2} = \phi_{1,1} + \delta$. The power ratio between the LoS and NLoS paths is set to be $20$ dB by simply setting $d_{1,2} = 10d_{1,1}$. The coordinates for the MS and the first R-RIS subarray are $\bp_{\text{MS}} = [0,0,0]^\mathsf{T}$ and $\bp_{\text{RIS},1} = [2, 4.6, 5.4]^\mathsf{T}$, respectively. The theoretical CRLB results for the effect of the value for $\delta$ on the estimation of $\theta_{1,1}$ and $\phi_{1,1}$ are shown in Fig.~\ref{Effect_of_NLoS_on_Ang_Est}, considering the parameter setting $K = 64$, $M_1 = 4 \times 4$, and $P = 0$ dBm. The analog combining matrix $\bW_1$ has been constructed as $[\bW_1]_{mn} = \exp(j[\boldsymbol{\Psi}]_{mn})$, where the random phase $[\boldsymbol{\Psi}]_{mn}$ follows $\mathcal{U}[0,2 \pi]$. The bandwidth $B$ is chosen to be $10$ MHz (so that the noise variance is $-174 + 10\log_{10}(10^6B) = -104$ dBm), and the carrier frequency is set to $28$ GHz. 

As depicted in Fig.~\ref{Effect_of_NLoS_on_Ang_Est}, when the deterministic angle difference $\delta$ increases within $[0,\pi/2]$, the performance of the LoS channel angular parameter estimation in the considered MPC scenario asymptotically approaches that in the pure LoS (i.e., single-path) scenario. This is illustrated in the figure by the two straight lines and the two curves marked with $\Delta = 0$. Note that $\Delta = 0$ stands for the case with deterministic and fixed angle difference $\delta$. The other $\Delta$ values in the figure will be introduced in the next example. The aforementioned observation is closely related to the correlation between the two subspaces spanned by $\bG_{1,1}$ and $\bG_{1,2}$. Two extreme cases are: \textit{i}) $\bG_{1,1} \approx 10\hat{\bG}_{1,2}$ with $\hat{\bG}_{1,2}\triangleq [\frac{\partial \boldsymbol{\mu}_1} {\partial \theta_{1,2}}, \frac{\partial \boldsymbol{\mu}_1} {\partial \phi_{1,2}} ]$, yielding $\bG_{1,1}^\mathsf{H} (\bI_K - \bP_{\bG_{1,2}})\bG_{1,1} \approx \mathbf{0}$; and \textit{ii}) $\bG_{1,1} \perp \bG_{1,2}$, where $\perp$ means being perpendicular/orthogonal (i.e., $\bG_{1,1}^\mathsf{H} \bG_{1,2} = \mathbf{0}$), which results in $ \bG_{1,1}^\mathsf{H} (\bI_K - \bP_{\bG_{1,2}})\bG_{1,1} = \bG_{1,1}^\mathsf{H} \bG_{1,1}$. The proofs for these two cases are delegated in the Appendix~\ref{Proofs}. These proofs indicate that when the angle difference increases, the performance on the angular parameters improves as well. When $\delta \approx 0$, case \textit{i}) applies, whereas when $\delta$ is large enough, case \textit{ii}) applies. As shown in Fig.~\ref{Effect_of_NLoS_on_Ang_Est}, when the angle difference is within the range $[0.6, 1.4]$ in radians, the performance of the LoS path's angular parameter estimation is approximately the same for both single-path and two-path scenarios.

\subsubsection{Example 2} Instead of relying on deterministic and fixed angle difference, in this example, we introduce randomness to the angle difference and evaluate its effect on the angular parameter estimation. We consider that the random angle difference $\hat{\delta}$ follows $\mathcal{U}[\delta - \Delta,  \delta + \Delta]$ with a fixed $\Delta$ value, which meets the following conditions: mean $\mathbb{E}\{\hat{\delta}\}= \delta$ and variance $\sigma_{\hat{\delta}}^2 = \Delta^2/3$. The other parameters are kept the same as those in \textit{Example 1}. The theoretical CRLB results are also included in Fig.~\ref{Effect_of_NLoS_on_Ang_Est}. As demonstrated, in the low $\delta$ regime, the effect of the randomness on angle difference, increasing proportionally to the variance of the angle difference (i.e., $\Delta^2/3$), is more significant compared to that in the high $\delta$ regime. 

\subsection{CRLB on 3D Localization}
The FIM of the MS's coordinates $\bp_{\text{MS}}$ can be obtained via the Jacobian matrix $\bT_i$, which is defined as $[\bT_i]_{mn} \triangleq \frac{\partial [\boldsymbol{\nu}_i]_n}{\partial [\bp_{\text{MS}}]_m}$ $\forall$$m=1,2,\ldots,3$ and $\forall$$n=1$ and $2$. Since we only exploit the azimuth and elevation AoAs of the LoS path in our 3D localization systems, i.e., $[ \boldsymbol{\nu}_i]_1$ and $[ \boldsymbol{\nu}_i]_2$, $\bT_i$ is a $3 \times 2$ matrix having the following elements~\cite{he2022simultaneous}:
\begin{align}\label{bT_i1}
    &[\bT_i]_{11} = \partial \theta_{i,1} / \partial  x_{\text{MS}}  =- \frac{\sin(\theta_{i,1})}{d_{i,1} \cos(\phi_{i,1})}, \\
    &[\bT_i]_{21} = \partial \theta_{i,1}/ \partial  y_{\text{MS}}  = \frac{\cos(\theta_{i,1})}{d_{i,1} \cos(\phi_{i,1})}, \\
    & [\bT_i]_{31} =\partial \theta_{i,1} / \partial  z_{\text{MS}}  = 0, \\
       & [\bT_i]_{12} = \partial \phi_{i,1} / \partial  x_{\text{MS}}  =- \frac{\cos(\theta_{i,1})\sin(\phi_{i,1})}{d_{i,1} },  \\
     &[\bT_i]_{22} =\partial \phi_{i,1}/ \partial  y_{\text{MS}}  =- \frac{\sin(\theta_{i,1})\sin(\phi_{i,1})}{d_{i,1} }, \\
     &[\bT_i]_{32} =\partial \phi_{i,1} / \partial  z_{\text{MS}}  = \frac{\cos(\phi_{i,1})}{d_{i,1}}.\label{bT_i6}
\end{align}
Hence, the FIM associated with the MS's coordinates, as estimated from the $i$th R-RIS subarray, can expressed as 
$\bJ(\bp_{\text{MS}}) = \bT_i  \bJ([\boldsymbol{\nu}_i]_{1:2}) \bT_i^\mathsf{T}$ where $\bJ([\boldsymbol{\nu}_i]_{1:2}) = \big(\bG_{i,1}^\mathsf{H} (\bI_{K} \!-\! \bP_{\bG_{i,2}})\bG_{i,1}\big)$. By adding all the contributions across the $I$ R-RIS subarrays, the Position Error Bound (PEB) of the proposed 3D localization system can be computed as follows: 
\begin{align}\label{PEB}
 &\sigma^2_{\hat{\bp}_\text{MS}} =  \mathbb{E}\{(\bp_\text{MS}  -\hat{\bp}_\text{MS})^\mathsf{T} (\bp_\text{MS} -\hat{\bp}_\text{MS})\}  \nonumber\\
   &\geq \mathrm{Tr}\Bigg\{\bigg(\sum_{i=1}^I \bT_i  \bJ([\boldsymbol{\nu}_i]_{1:2}) \bT_i^\mathsf{T}\bigg)^{-1}\Bigg\}, \nonumber \\
   & = \frac{\sigma_i^2}{2P} \mathrm{Tr}\Bigg\{\bigg(\sum_{i=1}^I \bT_i  (\bG_{i,1}^\mathsf{H} (\bI_{K} \!-\! \bP_{\bG_{i,2}})\bG_{i,1}) \bT_i^\mathsf{T}\bigg)^{-1}\Bigg\},
\end{align}
where $\hat{\bp}_\text{MS}$ represents the unbiased estimate of $\bp_\text{MS}$.

\section{Proposed 3D Localization}
In this section, we capitalize on the two-step theoretical analyses conducted in Section~\ref{CRLB_Analyses} and design a practical localization scheme. We first resort to the ANM and subspace-based root MUSIC algorithms to estimate the angular parameters for the LoS path at each R-RIS subarray, and then apply the LS principle to map those angular estimates to the 3D location of the MS. 

\subsection{Atomic Set and Norm}
ANM falls into the category of off-grid compressive sensing, which provides a powerful mathematical tool for eliminating the basis mismatch and power leakage problem, while offering high-accuracy angle estimation~\cite{Yang2015,Tang2013}. To this end, we start the definition of the atomic set~\cite{Yang2015, Yang2016,Tang2013, he2020anm} for our estimation problem:
\begin{equation}\label{atomic_set}
\mathcal{A} \triangleq \{\boldsymbol{\alpha}_y(x_1, x_2) \otimes \boldsymbol{\alpha}_z(x_2), x_1 \in [0, \pi], x_2 \in [-\pi/2, \pi/2]  \},  
\end{equation}
where each atom possesses the same structure with the linear terms in~\eqref{h_i}. For any vector $\bh_i \in \mathbb{C}^{M_{i,y} M_{i,z}}$ of the form $\bh_i = \sum_{l} \eta_l \boldsymbol{\alpha}_y(x_{1,l}, x_{2,l}) \otimes \boldsymbol{\alpha}_z(x_{2,l})$ with each $\eta_l>0$ being a coefficient, $x_{1,l} \in [0, \pi]$, and $x_{2,l} \in [-\pi/2, \pi/2]$, its atomic norm with respect to the atomic set $\mathcal{A}$ is written as follows:
\begin{align}
\|\bh_i\|_{\mathcal{A}} = &\mathrm{inf}_{\mathcal{B}}\Big\{\frac{1}{2M_{i,y} M_{i,z}} \mathrm{Tr}(\mathrm{Toep}(\mathcal{U}_2)) + \frac{t}{2}\Big\}, \nonumber\\
&\text{s.t.} \;\begin{bmatrix} \mathrm{Toep}(\mathcal{U}_2)  & \bh_i\\
\bh_i^{\mathsf{H}}& t_i
\end{bmatrix} \succeq \mathbf{0},
\end{align}
where set $\mathcal{B}\triangleq\{\mathcal{U}_2 \in \mathbb{C}^{M_{i,y} \times M_{i,z} }, t_i\in \mathbb{R}\}$ with $\mathcal{U}_2$ being a $2$-way tensor and $\mathrm{Toep}(\mathcal{U}_2)$ is a $2$-level block Toeplitz matrix, which results from the Vandermonde decomposition lemma for positive semidefinite Toeplitz matrices~\cite{Tang2013}.  

\subsection{Problem Formulation}
Based on the received signal model in~\eqref{by_i} and by taking into consideration the channel sparsity (in the form of $L_i \ll M_i$ $\forall i$) and the noise contribution, we consider the following regularized optimization problem formulation:
\begin{align}\label{ANM}
\min_{\bh_i \in \mathbb{C}^{M_i},\; \mathcal{B}} &\mu_i \|\bh_i\|_{\mathcal{A}} + \frac{1}{2}\|\by_i - \sqrt{P}\bW_i^\mathsf{H} \bh_i\|_2^2 \nonumber\\
&\text{s.t.} \;\begin{bmatrix} \mathrm{Toep}(\mathcal{U}_2)  & \bh_i\\
\bh_i^{\mathsf{H}}& t_i
\end{bmatrix} \succeq \mathbf{0},
\end{align}
where $\mu_i \propto \sigma_i \sqrt{ M_{i} \log(M_{i})}$ is the regularization term of the atomic norm penalty. This problem can be efficiently solved using the Matlab CVX toolbox~\cite{CVX}. Unlike the overcomplete dictionary, let alone the dictionary's undercomplete counterpart, in Orthogonal Matching Pursuit (OMP), since the number of atoms in~\eqref{atomic_set} is infinite, high accuracy recovery of $\bh_i$ can be achieved by well controlling the regularization term $\mu_i$. The computational complexity in this stage is $O\big(M_i^{3.5}\big)$~\cite{ZHANG201995}. After getting the estimate of $\bh_i$, denoted henceforth as $\hat{\bh}_i$, from~\eqref{ANM}, we can extract its angular parameters, e.g., via the subspace-oriented root MUSIC algorithm ~\cite{Barabell1983}, as detailed in the next subsection. 

\subsection{Azimuth and Elevation Angle Estimation}\label{sec:angle_estimation}
After obtaining $\hat{\bh}_i$, we formulate the estimated autocorrelation matrix $\hat{\bC}_i \triangleq \hat{\bh}_i \hat{\bh}_i^\mathsf{H} \in \mathbb{C}^{M_i  \times M_i}$. The ground-truth $\bC_i$ based on $\bh_i$ is expressed in~\eqref{bC_i_est} on the top of the next page,  
\begin{figure*}
\begin{align}\label{bC_i_est}
    \bC_i = \bh_i \bh_i^\mathsf{H} &= \left(\sum_{l = 1}^{L_i}\gamma_{i,l} \boldsymbol{\alpha}_y(\theta_{i,l},\phi_{i,l}) \otimes \boldsymbol{\alpha}_z(\phi_{i,l})\right) \left(\sum_{l = 1}^{L_i} \gamma_{i,l} \boldsymbol{\alpha}_y(\theta_{i,l},\phi_{i,l}) \otimes \boldsymbol{\alpha}_z(\phi_{i,l})\right)^\mathsf{H} \nonumber \\
    & = \sum_{l = 1}^{L_i} \sum_{k = 1}^{L_i}\gamma_{i,l}\gamma^*_{i,k} \underbrace{[\boldsymbol{\alpha}_y(\theta_{i,l},\phi_{i,l})  \boldsymbol{\alpha}_y(\theta_{i,k},\phi_{i,k})^\mathsf{H} ]}_{\bC^{(l,k)}_{i,y}}\otimes\underbrace{[  \boldsymbol{\alpha}_z(\phi_{i,l}) \boldsymbol{\alpha}_z(\phi_{i,k})^\mathsf{H}]}_{\bC^{(l,k)}_{i,z}},
\end{align}
\end{figure*}
where the following two properties of the Kronecker product have been considered: \textit{i}) $(\ba \otimes \bb)^\mathsf{H} = (\ba^\mathsf{H} \otimes \bb^\mathsf{H})$; and \textit{ii}) $(\ba \otimes \bb) (\bc \otimes \bd) = (\ba  \bc)\otimes(\bb  \bd)$. For simplicity, in~\eqref{bC_i_est}, we have introduced $\gamma_{i,l} \triangleq\frac{e^{-j 2\pi  d_{i,l}/\lambda}}{\sqrt{\rho_{i,l}}}$. Note that, from our assumptions in Section~\ref{sec_Sys_Model}, it holds that $|\gamma_{i,1}| \gg |\gamma_{i,2}| \geq \cdots \geq \gamma_{i,L_i}$. Finally, $\bC^{(l,k)}_{i,y} \in \mathbb{C}^{M_{i,y} \times M_{i,y}}$ and $\bC^{(l,k)}_{i,z}  \in \mathbb{C}^{M_{i,z} \times M_{i,z}}$ represent the cross-correlation matrices between $\boldsymbol{\alpha}_y(\theta_{i,l},\phi_{i,l})$ and $\boldsymbol{\alpha}_y(\theta_{i,k},\phi_{i,k})$ as well as between $\boldsymbol{\alpha}_z(\phi_{i,l})$ and $\boldsymbol{\alpha}_z(\phi_{i,k})$, respectively. We next write the ground-truth autocorrelation matrix $\bC_i$ in the following block matrix form:
\begin{equation}\label{bC_i}
      \bC_i  = 
      \begin{bmatrix}
     [\bC_{i}]_{11}  & * & *\\
     *&\ddots&* \\
      * & *& [\bC_{i}]_{M_{i,y} M_{i,y}} 
      \end{bmatrix}.
\end{equation}
By following~\eqref{alpha_y} and \eqref{bC_i}, the diagonal elements of $\bC_{i}$ with $m=1,2,\ldots,M_i$ are given by
\begin{align}\label{bC_i_z}
    [\bC_{i}]_{mm}
    &= \sum_{l = 1}^{L_i} \sum_{k = 1}^{L_i}\gamma_{i,l}\gamma^*_{i,k} e^{j\pi (m-1)  \varphi_{i,l,k}} [  \boldsymbol{\alpha}_z(\phi_{i,l}) \boldsymbol{\alpha}_z(\phi_{i,k})^\mathsf{H}], \nonumber \\
    & = \Big[\sum_{l = 1}^{L_i}e^{j\pi (m-1) \sin(\theta_{i,l}) \sin(\phi_{i,l})}\gamma_{i,l}\boldsymbol{\alpha}_z(\phi_{i,l})\Big] \nonumber\\
    &\hspace{0.45cm}\times \Big[\sum_{l = 1}^{L_i}e^{j\pi (m-1) \sin(\theta_{i,k}) \sin(\phi_{i,k})}\gamma_{i,k}\boldsymbol{\alpha}_z(\phi_{i,k})\Big]^\mathsf{H},
\end{align}
where $\varphi_{i,l,k} \triangleq\sin(\theta_{i,l}) \sin(\phi_{i,l}) - \sin(\theta_{i,k}) \sin(\phi_{i,k})$. We know, however, that the submatrices included on the diagonal of $\bC_i$ are related to the autocorrelation matrix of $\sum_{l = 1}^{L_i}e^{j\pi (m-1) \sin(\theta_{i,l}) \sin(\phi_{i,l})}\gamma_{i,l}\boldsymbol{\alpha}_z(\phi_{i,l})$, which is the linear combination of the array response vectors $\boldsymbol{\alpha}_z(\phi_{i,l})$ for $l = 1,2,\ldots,L_i$.  We calculate $\bC_{i,z} = \sum_{m = 1}^{M_{i,y}} [\bC_{i}]_{mm}$, based on which
we can apply the root MUSIC algorithm~\cite{Barabell1983} for estimating $\phi_{i,1}, \ldots, \phi_{i,L_i}$.\footnote{In this paper, we assume that the gain of the LoS path is much larger than that of any of the NLoS paths. In this way, it is highly probable to find the correct LoS AoA without encountering any path association/mismatch issue.} In practice, we replace $\bC_{i}$ with $\hat{\bC}_{i}$ for performing the estimation. The calculation of
~\eqref{bC_i_z} involves $M_{i,y} L_i^2 M^2_{i,z}$ addition operations and $M_{i,y} L_i^2 M^2_{i,z}$ multiplication operations. In addition, the computational complexity for root MUSIC here is $O\big(M_{i,z}^{3}\big)$.

Similarly, based on the definition of the Kronecker product $[\bC^{(l,k)}_{i,y} \otimes \bC^{(l,k)}_{i,z} ]_{M_{i,z} (r-1) +m \;\; M_{i,z} (s-1) +n  } = [\bC^{(l,k)}_{i,y} ]_{rs} [\bC^{(l,k)}_{i,z} ]_{mn}$, we can extract $\bC^{(l,k)}_{i,y}$ from $\bC_i$ corresponding to  $[\bC^{(l,k)}_{i,z} ]_{m m} = e^{j\pi (m-1)  [\cos(\phi_{i,l}) - \cos(\phi_{i,k})] } $. By doing this, the autocorrelation matrix $\bC_{i,y}^{(m)}$ for $\sum_{l = 1}^{L_i}e^{j\pi (m-1) \cos(\phi_{i,l})}\gamma_{i,l}\boldsymbol{\alpha}_y(\theta_{i,l},\phi_{i,l})$ is obtained, which is the linear combination of the array response vectors $\boldsymbol{\alpha}_y(\theta_{i,l},\phi_{i,l})$ for $l = 1,2,\ldots,L_i$. Similarly, after getting $\bC_{i,y} = \sum_{m=1}^{M_{i,z}}\bC_{i,y}^{(m)}$, we can extract the estimation of $\theta_{i,1}$ using via the root MUSIC algorithm~\cite{Barabell1983}, which, as previously mentioned, has computational complexity $O\big(M_{i,y}^{3}\big)$.

\subsection{Estimation of the 3D Location Coordinates}
After estimating the AoAs of the LoS paths at each $i$th R-RIS subarray (i.e., $\theta_{i,1}$ and $\phi_{i,1}$ $\forall i = 1,2,\ldots,I$), we apply the LS principle for mapping those estimates to the 3D position of MS, $\bp_{\text{MS}}$, as follows~\cite{Alexandropoulos2022}:
\begin{equation}\label{LS_Loc}
    \hat{\bp}_{\text{MS}} = \left(\sum_{i = 1}^I \bB_i\right)^{-1} \left(\sum_{i = 1}^I \bB_i \bp_{\text{RIS},i} \right),
\end{equation}
where $\bB_i \triangleq \bI_3 - \hat{\boldsymbol{\xi}}_i \hat{\boldsymbol{\xi}}_i^\mathsf{T}$ and $\hat{\boldsymbol{\xi}}_i \triangleq [\cos(\hat{\theta}_{i,1}) \cos(\hat{\phi}_{i,1})  , \sin(\hat{\theta}_{i,1}) \cos(\hat{\phi}_{i,1}), \sin(\hat{\phi}_{i,1})   ]^\mathsf{T}$. We henceforth use the symbols $\hat{\theta}_{i,1}$ and $\hat{\phi}_{i,1}$ for the estimates of the azimuth and elevation AoAs, respectively, associated with the LoS path in each $\bh_i$. The computational complexity of this operation can be neglected due to the small size of $\bB_i$.

In order to eliminate the negative effect introduced by potentially poor angular estimates, we resort to an \textit{outlier finding} approach to find out any outliers and exclude their contributions in~\eqref{LS_Loc}. The outliers are values which are by default more than three times of the Median Absolute Deviations (MAD) away from the median~\cite{leys2013detecting}. We specifically seek for any outliers in the sets $\{\hat{\theta}_{1,1},\ldots,\hat{\theta}_{I,1}\}$ and $\{\hat{\phi}_{1,1},\ldots,\hat{\phi}_{I,1}\}$ separately. If neither $\hat{\theta}_{i,1}$ nor $\hat{\phi}_{i,1}$ is an outlier, we include them in~\eqref{LS_Loc}; otherwise, their contributions are excluded. 

Putting all above together, the overall computational complexity of the proposed 3D localization approach is dominated by the ANM technique, which brings $O\big(M_i^{3.5}\big)$ computational complexity. The major steps of the proposed 3D localization scheme are summarized in \textbf{Algorithm~\ref{alg:3d_loc}}.

\begin{algorithm}[t]
\caption{Proposed 3D Localization}\label{alg:3d_loc}
 \hspace*{\algorithmicindent} \textbf{Input}: $\by_i$ from~\eqref{by_i}, $\bW_i$, $P$, $M_i$, and $\sigma_i$.\\
 \hspace*{\algorithmicindent} \textbf{Output}: $\hat{\bp}_{\text{MS}}$ via~\eqref{LS_Loc}; 
\begin{algorithmic}[1]
\State Solve the formulated ANM problem in~\eqref{ANM} to get $\hat{\bh}_i$;
\State Decompose the autocorrelation matrix $\hat{\bC}_i \triangleq \hat{\bh}_i \hat{\bh}_i^\mathsf{H}$ into $\bC_{i,z}$ and $\bC_{i,y}$, as described in~Section~\ref{sec:angle_estimation}; 
\State Estimate the azimuth and elevation AoAs from $\bC_{i,z}$ and $\bC_{i,y}$ via the root MUSIC algorithm;
\State Map the angular estimates to coordinates of the targeted MS using~\eqref{LS_Loc}.
\end{algorithmic}
\end{algorithm}

\section{Numerical Results}\label{Numerical_results}
In this section, we investigate the feasibility of the proposed single R-RIS-based 3D localization approach, which is based on the R-RIS partitioning into several single-RX-RF R-RIS subarrays. In the performance evaluation results that follow, without otherwise stated, the positions of the MS and $I=4$ R-RIS subarrays are set as: $\bp_{\text{MS}} = [0,0,0]^\mathsf{T}$, $\bp_{\text{RIS},1} = [2, 4.6, 5.4]^\mathsf{T}$, $\bp_{\text{RIS},2} = [2, 4.6, 4.6 ]^\mathsf{T}$, $\bp_{\text{RIS},3} = [2, 5.4, 5.4]^\mathsf{T}$, and  $\bp_{\text{RIS},4} = [2, 5.4, 4.6]^\mathsf{T}$, where the centroid of the overall R-RIS is set to $\bp_{\text{RIS,c}} = [2, 5, 5]^\mathsf{T}$. By following the parameter setup of \textit{Example 1} in Section~\ref{Effect_of_NLoS_Paths}, each R-RIS subarray has $16 = (4 \times 4)$ meta-atom elements. The bandwidth $B$ is chosen to be $10$ MHz and the carrier frequency is set to $28$ GHz. The analog combining matrix in~\eqref{by_i} is constructed from the first $K$ columns of a Discrete Fourier Transform (DFT) matrix; note that similar trends are expected from practical RIS phase configuration codebooks~\cite{Rahal_2022}. All individual channels in the results that follow have been generated using the general model~\eqref{NF_steering_vector}. The number of paths for each R-RIS subarray channel is set as $L_i = 2$,\footnote{It is noted that the proposed localization algorithm can be directly applied to cases with $L_i >2$.} where the average power ratio between LoS and NLoS paths was set to $20$ dB. All the simulations results are obtained via $1000$ trials. The parameters setup is summarized in Table~\ref{tab1:parameter}.

\begin{table}[t]
    \centering
  \caption{Parameters' Setup in the Simulation Results.}
    \label{tab1:parameter}
    \begin{tabular}{cc|cc}
        \hline
        Parameter  & Value &Parameter  & Value  \\
      \hline
        $M_i$ & $16$  & $\delta$ & $\pi/4$ \\
        $I$ & $4$ & $\Delta$ & $0$ \\
        $K$&$\{16, 32, 64\}$& $B$ & $10$ MHz \\
        $L_i$ & $2$ & $f_c$ & $28$ GHz \\
       $\bp_\text{MS}$&$[0, 0, 0]^\mathsf{T}$&  $\bp_{\text{RIS},1}$  &  $[2, 4.6, 5.4]^\mathsf{T}$   \\
        $\bp_{\text{RIS},2}$  &  $[2, 4.6, 4.6 ]^\mathsf{T}$ &
       $\bp_{\text{RIS},3}$  &  $[2, 5.4, 5.4]^\mathsf{T}$  \\
       $\bp_{\text{RIS},4}$  &  $[2, 5.4, 4.6]^\mathsf{T}$  &
        $\bp_{\text{RIS,c}} $ & $[2, 5, 5]^\mathsf{T}$  \\
      \hline
    \end{tabular}
\end{table}

\subsection{Effect of the Training Overhead}\label{subsec_training_overhead}
\begin{figure}[t]
	\centering
\includegraphics[width=0.99\linewidth]{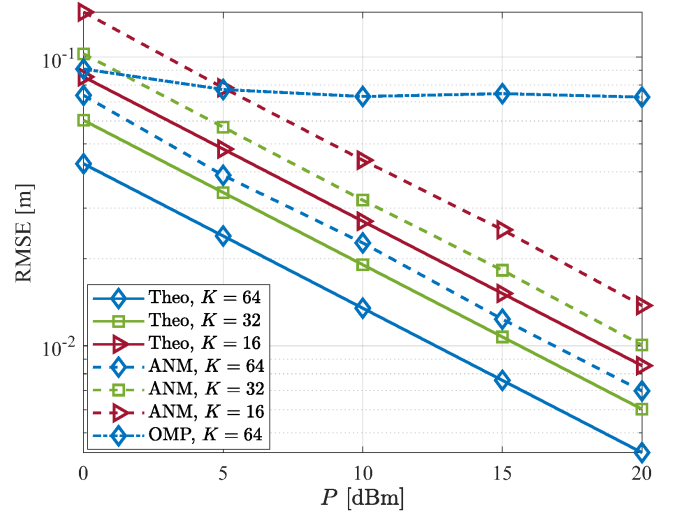}
	\caption{RMSE performance of the proposed 3D localization system with different training overhead values.}
		\label{Training_overhead_Delta_dot25pi_theo_prac}
\end{figure}

The simulation results in terms of the Root Mean Square Error (RMSE) in meters, i.e., $\sigma_{\hat{\bp}_\text{MS}}$ in~\eqref{PEB}, as a function of the transmit power $P$ in [dBm] are included in Fig.~\ref{Training_overhead_Delta_dot25pi_theo_prac}, where different training overheads values $K = \{16, 32, 64\}$ and the fixed $\delta = \pi/4$ ($\Delta = 0$) are considered. Results obtained from the OMP algorithm considering very high resolution on the quantization of azimuth and elevation angles (we specifically used $2048$ grid points for each dimension) are also included for performance comparison. As shown in the figure and as expected, the higher the training overhead is, the better becomes the estimation performance. When $P = 20$ dBm, ANM with $K = 64$ can achieve roughly $0.7$ cm localization accuracy, while with $K = 16$ the accuracy is around $1$ cm. Taking into consideration LoS information can bring cm-level localization accuracy. Thus, there is no need to exploit NLoS path information, which will significantly increase the computational complexity of the algorithm. The figure also depicts that the performance of the introduced OMP saturates to $7$ cm when the transmit power increases due to inevitable quantization error; recall that, for the proposed algorithm, an extremely large dictionary has been used. It is also evident in Fig.~\ref{Training_overhead_Delta_dot25pi_theo_prac} that the performance gap between the theoretical and practical results stays constant within the studied SNR range. Under the same RMSE level, the gap is roughly $5$ dB in terms of the transmit power.

\subsection{Effect of the Distance between the MS and the R-RIS}\label{Effect_distance_MS_BS}
\begin{figure}[t]
	\centering
\includegraphics[width=0.99\linewidth]{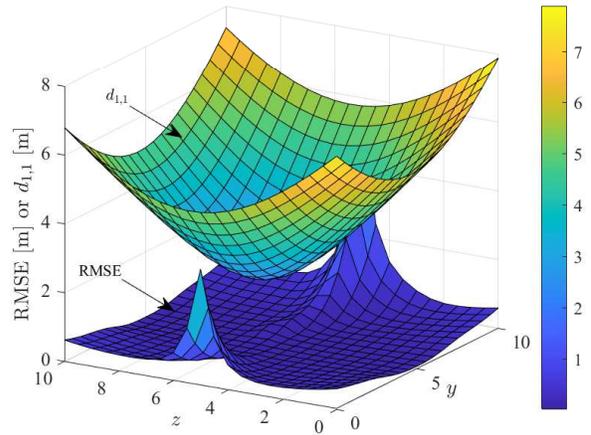}
	\caption{RMSE performance of the proposed localization system with different MS-to-R-RIS distances $d_{1,1}$. Recall from our assumptions that $d_{1,1} \approx d_{2,1} \approx d_{3,1} \approx d_{4,1}$. }
		\label{Heat_map1}
\end{figure}
In this experiment, we evaluate the effect of MS-to-R-RIS distance on the RMSE performance, where we fix the $x$-coordinate of the MS to be $0$, while changing the other two coordinates. We set $P = -20$ dBm and $K = 32$ and keep the other system parameters as in Table~\ref{tab1:parameter}. The simulation results are shown in Fig.~\ref{Heat_map1}. Note that we only include $d_{1,1}$ in the figure since the other distances $d_{2,1},\ldots,d_{4,1}$ have been assumed to be approximately the same. As can be observed from the figure, the performance is not strictly inversely proportional to the distance, since the angle values also play an important role in the 3D localization performance. 

Let's assume that $\bJ^{-1}([\boldsymbol{\nu}_i]_{1:2}) \geq \mathrm{diag}([\epsilon_{i,1}, \epsilon_{i,2}]^\mathsf{T} )$, where $\epsilon_{i,1}$ and $\epsilon_{i,2}$ are introduced to characterize the performance bounds on the angular parameter estimation related to the LoS path. The FIM for the 3D coordinates is $ \bJ(\bp_\text{MS}) = \sum_{m =1}^2\epsilon_{i,m}^{-1} \bt_{i,m}\bt_{i,m}^\mathsf{T}$, where $\bt_{i,m}$ is the $m$th column vector of the Jacobian matrix $\bT_i$. According to \eqref{bT_i1}--\eqref{bT_i6}, it holds that $\bt_{i,m}$'s are orthogonal vectors. Therefore, $\mathrm{tr}\{ \bJ^{-1}(\bp_\text{MS}) \} = \sum_{m =1}^2 \epsilon_{i,m}/ \|\bt_{i,m}\|_2^2 = \epsilon_{i,1} d_{i,1}^2 \cos^2(\phi_{i,1}) + \epsilon_{i,2} d_{i,1}^2 $. It can be concluded from this expression that the position error variance increases linearly with the angular parameter estimation error variance and quadratically with the distance between the MS and the $i$th R-RIS subarray. We also note that, when $d_{i,1}\gg 1$, the requirement for high accuracy of the angular parameter increases if we aim at achieving the same level of positioning accuracy as for a case with a small $d_{i,1}$. We further know that the 3D localization not only depends on the angular parameter estimation, but also on the elevation angle and distance. These factors jointly affect the 3D localization accuracy, as revealed in Fig.~\ref{Heat_map1}. It can be concluded from this investigation that, when $z\approx 5$ results in $\phi_{i,1} \approx 0$, which constitutes the worst case for our 3D localization system, since satisfying performance becomes infeasible. 

\subsection{Effect of the Inter R-RIS Subarray Spacing}
\begin{figure}[t]
	\centering
\includegraphics[width=0.99\linewidth]{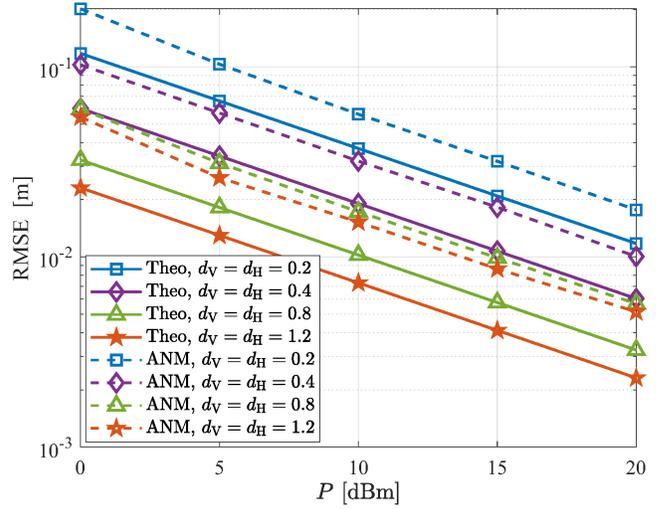}
	\caption{Effect of the inter R-RIS subarray spacing on the RMSE performance from both the theoretical and practical perspectives.}
		\label{Inter_sub_RIS_spacing}
\end{figure}

We hereinafter consider different inter R-RIS subarray spacings and evaluate the their impact on the 3D localization accuracy. We assume that the centroid of the whole R-RIS is located at $[2,5,5]$ and change the vertical and horizontal distances of the R-RIS subarrays to it. For instance, in Sections~\ref{subsec_training_overhead} and~\ref{Effect_distance_MS_BS}, both the vertical and horizontal distances $d_\text{V}$ and $d_\text{H}$ are set to $0.4$, i.e., $d_\text{V} = d_\text{H} = 0.4$. We extend it to the cases where $d_\text{V} = d_\text{H} = 0.8$, $d_\text{V} = d_\text{H} = 1.2$, and $d_\text{V} = d_\text{H} = 0.2$. The simulation results are shown in Fig.~\ref{Inter_sub_RIS_spacing} with $K = 32$. As can be seen from the figure, when fixing $P = 10$ dBm, the case where $d_\text{V} = d_\text{H} = 1.2$ with the proposed localization scheme can achieve around $2$-cm accuracy, while when $d_\text{V} = d_\text{H} = 0.2$ only $6$-cm accuracy can be reached. As a conclusion, larger spacing in general can bring better localization performance, but at a sacrifice of a larger space for installing the overall R-RIS structure.  

\subsection{R-RIS Partitioning Patterns}\label{subsec_RIS_partitioning}
In addition to the $2\times 2$ R-RIS partitioning pattern studied so far, we further introduce another two partitioning designs for the R-RIS: \textit{i}) all the R-RIS subarrays are deployed vertically in a line, i.e., $1\times 4$; and \textit{ii}) all the R-RIS subarrays are deployed horizontally in a line, i.e., $4 \times 1$. The rest of the parameters are set as follows: $d_\text{V} = 0.4$ and/or $d_\text{H} = 0.4$, and $K = 32$. In order to examine these designs, we introduce the concept of the Geometric Dilution of Precision (GDoP)~\cite{Moragrega2013}, expressed in the form of $\sqrt{\mathrm{Tr}\{(\bH^\mathsf{T}\bH)^{-1}\}}$, where $\bH = [\bp_{\text{RIS},1}/d_{1,1}; \bp_{\text{RIS},2}/d_{2,1};\bp_{\text{RIS},3}/d_{3,1};\bp_{\text{RIS},4}/d_{4,1}]^\mathsf{T}\in \mathbb{R}^{4\times 3}$ with the semicolons implying the separation of the column vectors in the matrix. For the calculation of the GDoP value, we replace the matrix inverse with the Moore–Penrose on in the previous expression, whenever $\bH^\mathsf{T}\bH$ is singular. It is noted that, in general, holds that the smaller the GDoP value is, the better is the deployment of the anchors. 

\begin{figure}[t]
	\centering
\includegraphics[width=1\linewidth]{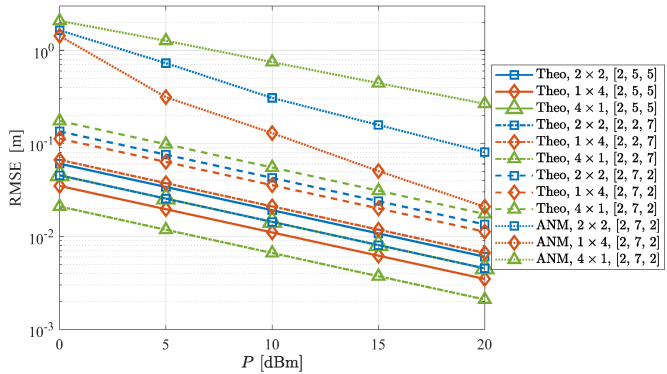}
	\caption{Effect of R-RIS partitioning on the RMSE performance from both the theoretical and practical perspectives.}
		\label{RIS_Partitioning_pattern}
\end{figure}

The simulation results on 3D localization are provided in Fig.~\ref{RIS_Partitioning_pattern} considering different R-RIS centroids, as marked in the legend. Note that we only provide simulation results for the cases with centroid $[2,7,2]$, consistent with theoretical limits, in order to avoid overcrowding of the figure. The GDoP values and theoretical RMSE (via CRLB) values at $P = 20$ dBm for all the partitioning patterns are included in Table~\ref{GDoP}, where we mark the RMSE values from the proposed 3D localization scheme in bold for the last three cases. It can be seen from the figure that, for most of the cases, the vertical deployment of the R-RIS subarrays yields the best performance, e.g., the patterns ``$1\times 4, [2,5,5]$'' and ``$1\times 4, [2,7,2]$'' outperforming the other two cases with the same R-RIS centroid, which is consistent with the GDoP analyses in Table~\ref{GDoP}. Note that the GDoP provides a rule of thumb regarding the ``anchors'' geometry for localization purposes. In other words, it is independent from the localization algorithm itself, but it gives a rough idea about the anchors deployment (i.e., the R-RIS subarrays deployment in this paper). In terms of the trade-off between compactness and performance, the $2 \times 2$ R-RIS partitioning is shown to stand out among the others considered in Fig.~\ref{RIS_Partitioning_pattern}. 
\begin{table}[t]
    \centering
  \caption{GDoP values and theoretical RMSE (via CRLB) values in meters at $P = 20$ dBm for different R-RIS partitioning patterns and R-RIS centroids.}
    \label{GDoP}
    \begin{tabular}{ccc}
        \hline
        Design  & GDoP Value & RMSE at $P = 20$ dBm  \\
      \hline
        $2\times 2, [2,5,5]$ & $34.7729$  &  $0.0060$\\
        $1\times 4, [2,5,5]$ & $5.5910$  &$0.0035$\\
        $4\times 1, [2,5,5]$ & $5.5910$  & $0.0045$\\
        $2\times 2, [2,2,7]$ & $36.7671$ &$0.0045$\\
        $1\times 4, [2,2,7]$ & $11.0305$&$0.0066$ \\
        $4\times 1, [2,2,7]$ & $4.4438$ & $0.0021$\\
        $2\times 2, [2,7,2]$ & $36.7671$&$0.0134\; \mathbf{(0.0804)}$ \\
        $1\times 4, [2,7,2]$ & $4.4438$&$0.0112\;\mathbf{(0.0207)}$ \\
       $4\times 1, [2,7,2]$ & $11.0305$ & $0.0174\; \mathbf{(0.2664)}$\\
      \hline
    \end{tabular}
\end{table}

\section{Conclusions}
In this paper, we have introduced the R-RIS that is capable to receive pilot signals and perform 3D localization of the MSs transmitting them, in a cost-efficient manner. This operation is accomplished via a few subarrays of meta-atoms hardware architecture, where each subarray is attached to a reception RF chain, enabling estimation of the impinging signal's AoAs. By using the spatially sampled versions of the received pilots, we have presented a localization method based on off-grid ANM and root MUSIC for the per subarray angular parameter estimates, which was followed by a fused LS approach for computing the MS's 3D coordinate estimates. Our extensive performance evaluation, which has been verified by our derived CRLBs for the estimation parameters, demonstrated the impact of the number of pilots, MS-to-R-RIS distance, inter R-RIS subarray spacing, and the R-RIS partitioning pattern on the localization performance.

The proposed 3D localization system, which does not require any access point and operates in far-field conditions of the R-RIS subarrays, has the following advantageous features: \textit{i}) it does not require ultra-wide bandwidth, relying only on AoA estimates; \textit{ii}) intermediate AoA estimates possess similar performance thanks to the close proximity among the R-RIS subarrays, rendering the presented LS-based mapping of the AoAs to the MS 3D Cartesian coordinate sufficient; and \textit{iii}) it provides accurate MS location estimation with three reception RF chains. For future work, we intend to  extend the considered framework to tracking of mobile multi-antenna MSs, and study the optimization of the R-RIS partitioning pattern in subarrays of meta-atoms.  


\section{Acknowledgement}
The work of Profs.~G.~C.~Alexandropoulos and H.~Wymeersch has been supported by the EU H2020 RISE-6G project under grant number 101017011.

 \appendix
\section{Proofs} \label{Proofs}

Let $\mathcal{R}(\bG_{1,m})$ be the column space of $\bG_{1,m}$ defined in Section~\ref{Effect_of_NLoS_Paths}. Then, for any matrix $\bM$ such that $\mathcal{R}(\bM) \subset \mathcal{R}(\bG_{1,2})$, the equality $\bP_{\bG_{1,2}} \bM = \bM$ holds, since $\bP_{\bG_{1,2}}$ is the projector onto $\mathcal{R}(\bG_{1,2})$.
Assume first the strict equality in the case \textit{i}), namely $\bG_{1,1} = 10\hat{\bG}_{1,2}$. It follows that $\mathcal{R}(\bG_{1,1}) \subset \mathcal{R}(\bG_{1,2})$ yielding either $\bP_{\bG_{1,2}} \bG_{1,1} = \bG_{1,1}$ or $(\bP_{\bG_{1,2}} - \bI_K) \bG_{1,1} = \mathbf{0}$. Multiplying $\bG_{1,1}^{\mathsf{H}}$ on the left completes the proof. For the case \textit{ii}), the hypothesis that $\bG_{1,1} \perp \bG_{1,2}$ means that $\mathcal{R}(\bG_{1,1})$ is in the left null space of $\bG_{1,2}$. Since the projector onto the left null space of $\bG_{1,2}$ is  $\bI_K - \bP_{\bG_{1,2}}$, the equality $(\bI_K -~\bP_{\bG_{1,2}})\bG_{1,1} = \bG_{1,1}$ holds. Multiplying $\bG_{1,1}^{\mathsf{H}}$ on the left completes the proof.

\bibliographystyle{IEEEtran}
\bibliography{IEEEabrv,Ref}

\end{document}